# Hybrid density functional theory LCAO calculations on phonons in Ba (Ti,Zr,Hf) O$_3$

Robert A. Evarestov *

*Department of Quantum Chemistry, St. Petersburg University, Universitetsky Prosp. 26, 198504 St. Petergof, Russia*

Phonon frequencies at Γ,X,M,R-points of Brilloin zone in cubic phase of Ba(Ti,Zr,Hf)O$_3$ were first time calculated by frozen phonon method using density functional theory (DFT) with hybrid exchange correlation functional PBE0. The calculations use linear combination of atomic orbitals (LCAO) basis functions as implemented in CRYSTAL09 computer code. The Powell algorithm was applied for basis set optimization. In agreement with the experimental observations the structural instability via soft modes was found only in BaTiO$_3$. A good quantitative agreement was found between the theoretical and experimental phonon frequency predictions in BaTiO$_3$ and BaZrO$_3$. It is concluded that the hybrid PBE0 functional is able to predict correctly the structural stability and phonon properties in perovskites.

Barium and Hafnium zirconates (BaZrO$_3$-BZO, BaHfO$_3$-BHO) have a perfect cubic structure ( space group Pm3m) over a wide range of temperatures and do not undergo any structural phase transitions even down to 2K.  Barium titanate (BaTiO$_3$-BTO) has  the cubic structure   only at high temperatures.  At 393 K BTO undergoes a structural phase transition to a tetragonal ferroelectric phase, below 278 K and 183 K BTO has an orthorhombic and rhombohedral structures, respectively [1].

The first principles Full-Potential LAPW [2-3] and Plane Waves (PW) DFT [4-9] phonon spectra calculations on BTO cubic phase demonstrated its structural instability via soft modes with imaginary frequencies at the BZ symmetry points Γ (0, 0, 0), X(0, 0, 0.5) and M(0,0.5,0.5).

The first principles DFT calculations on phonons in BZO give contradictory results [10]. Indeed the imaginary frequencies (instable modes) at Γ point were not found in DFT calculations [11-14] but appear in the results of calculations [10]. The instable mode at M was not  found in [11-12] but appears in the results of calculations [10] performed using also DFT method. The instable mode at R was found in calculations [10-12] in disagreement with the experimental data that BZO structure is cubic. It was concluded [10] that the examination of phonon instabilities and their consequences for structure symmetry is especially demanding of the quality of the pseudopotential being used: transferable small-core pseudopotentials (transferable rather than inexpensive) should be taken.

To the best of our knowledge only one phonon DFT (LDA) calculation is published [15] for cubic BHO and there were not found imaginary frequencies, in agreement with the experimentally found stability of cubic BHO.

In the present study the same periodic LCAO approach is applied to calculate the ground state properties and phonon frequencies for all the three crystals Ba (Ti,Zr,Hf) O$_3$ within the hybrid exchange-correlation functional, namely PBE0[16].

The hybrid PBE0 functional does not require any adjustable parameters and has been deduced on the basis of pure theoretical grounds. This fun ctional has shown to properly re-produce the properties of many strongly correlated systems (see, for example, Ref. 17).

The results obtained are compared with the existing experimental data and with those from PW DFT calculations.

The cubic perovskites properties and phonon frequencies in the LCAO basis were calculated in Crystal09[18]. The 5s, 5p and 6s levels of barium, the ns, np, (n+1)d, (n+1)s levels of transition metal were treated as valence states (n=3,4,5 for Ti, Zr, Hf, respectively ).The Gaussian basis set and relativistic pseudopotentials (PP) of free atoms for Zr and Hf were taken from the PPs library of the Stuttgart/Cologne group[19]. The uncontracted Gaussian basis set and small-core pseudopotential for Ti atom was taken from [20]. In order to avoid spurious interactions between the diffuse functions and the core functions of neighboring atoms, the basis set diffuse exponents smaller than 0.1 (in Bohr$^{-2}$) for Ti, Zr and Hf were removed. The all electron basis set 8-411G* from Ref. [21] was used for O atom. The exponents of non-contracted basis functions were optimized by the method of conjugate directions[22] as implemented in the OPTBAS code[23]. The Monkhorst-Pack scheme for 8x8x8 k-point mesh in the Brillouin zone (BZ) was applied together with the tolerances 7, 7, 7, 7, 14 for the Coulomb and exchange integrals calculations. Furthermore, the forces for the self consistent cycles were optimized until the energy difference reached $10^{-6}$ eV for the lattice structure optimization and $10^{-8}$ eV for the phonon frequency calculations.

Table 1 presents the results of our calculations for structure parameter (lattice constant *a*), bulk modulus *B*, cohesive energy $E_{at}$, the electronic band gap $E_{gap}$, effective charges (due to Mulliken population analysis in CRYSTAL09[24] and Born atomic charges). The results are also compared to existing experimental data. As it is seen from Table 1 use of the hybrid Hartree-Fock-DFT Hamiltonian allows predicting cubic perovskites properties with high accuracy.

It is worth mentioning that the standard DFT exchange-correlation functional (LDA,GGA) underestimates the band gap and overestimates cohesive energy[24].

**Table 1.** Calculated bulk properties of cubic $BaTiO_3$, $BaZrO_3$, $BaHfO_3$ crystals. The experimental data are given in parenthesis. For Born charges in $BaTiO_3$ experimental data are taken from [35].

| Property | $BaTiO_3$ * | $BaZrO_3$ | $BaHfO_3$ |
|---|---|---|---|
| Cell $a$, Å | 3.988 (3.992) | 4.198 (4.191,[26]) | 4.193 (4.180, [29]) |
| Bulk modulus $B$, GPa | 185 (173) | 168 | 171 |
| Cohesive energy $E_{at}$, eV | 29.9 (31.6) | 32.9 (33.5, [27]) | 33.4 |
| Band gap, $E_{gap}$, eV | 3.7 (3.4) | 5.4 (5.3, [28]) | 5.9 (6.0[15])** |
| $Q_{Ba}$ | 1.90 | 1.87 | 1.87 |
| $B_{Ba}$ | 2.60 (2.9) | 2.56 | 2.55 |
| $Q_{Me}$ | 2.24 | 2.21 | 2.48 |
| $B_{Me}$ | 7.06 (6.7) | 5.88 | 5.56 |
| $Q_O$ | -1.38 | -1.36 | -1.45 |
| $B_{O1}$ | -5.64 (-4.8) | -4.58 | -4.28 |
| $B_{O2}$ | -2.01(-2.4) | -1.94 | -1.91 |

*The experimental data are given as cited in [25] **
**Predicted from experimental band gap for $SrHfO_3$

The calculations of phonon frequencies were based on the following procedure. First, the equilibrium geometry was found by optimizing the structure parameter $a$. Second, the phonon frequencies were obtained within the frozen phonon method[30-31] within the harmonic approximation at the optimized equilibrium lattice constant. A dynamical matrix was calculated for the supercells containing 8, 32 and 64 primitive cells (40,160,320 atoms). The primitive cell - supercell transformation matrixes are $\begin{bmatrix} 200 \\ 020 \\ 002 \end{bmatrix}, \begin{bmatrix} 22-2 \\ 2-22 \\ -222 \end{bmatrix}, \begin{bmatrix} 400 \\ 040 \\ 004 \end{bmatrix}$ respectively. All these supercells are compatible with $\Gamma$,X,M,R - points of BZ ( at these points the difference in the calculated for 3 supercells frequencies is not larger than 1-2cm$^{-1}$). Two larger supercells are compatible also with **k**-points on the BZ directions: $\Sigma$(0.25,0.25,0),S(0.25,0.25,0.5)-for both of them and $\Delta$(0,0,0.25), T(0.25,0.5,0.5), $\Lambda$(0.25,0.25,0.25),Z(0,0.25,0.5) - for the largest supercell.

The phonon mode symmetry is defined by the corresponding space group irreducbile representations (irreps) induced from atom site symmetry group irreps corresponding to the atomic displacement x,y,z. [24]. In the cubic perovskite structure barium and transition metal atoms occupy Wyckoff positions a and b with the site symmetry group $O_h$ and oxygen atom occupies Wyckoff position d with the site symmetry $D_{4h}$. In footnote to Table 2 the symmetry of phonons is given for $\Gamma$ point of BZ. It is seen that IR active $\Gamma_4^-$ modes are induced by all the five atoms displacements, but the silent $\Gamma_5^-$ mode- only by three oxygens displacements. The symmetry of phonons for X,M,R points of BZ is given in Table 3 with the corresponding frequencies. The irreducible representations of the cubic space group Pm3m are labeled according to Miller and Love [32].

Phonon frequencies at $\Gamma$ point of Brillouin zone are given in Table 2 in comparison with those found in PW DFT calculations by linear response (not frozen phonon!) method. It is seen that: 1) in agreement with the results of other calculations and experimental observations the structural instability via soft modes was found only in $BaTiO_3$; 2) the best agreement between the experimental and calculated frequencies takes place for the present LCAO PBE0 calculations.

Table 2 Phonon frequencies (cm$^{-1}$) at $\Gamma$ point of Brillouin zone

| phonon | $BaTiO_3$ this work | [4] | [9] | [5] | [6] | **exper. [33]** | $BaZrO_3$ this work | [12] | [13] | [14] | **exper. [34]** | $BaHfO_3$* this work | [15] |
|---|---|---|---|---|---|---|---|---|---|---|---|---|---|
| 4$^-$ TO1 | 273i | 178i | 151i | 195i | 177i | - | 120 | 96 | 91 | - | **115** | 118 | 94 |
| 4$^-$ TO2 | 194 | 177 | 175 | 166 | 112 | **181** | 220 | 193 | 190 | 250 | **210** | 218 | 194 |
| 5$^-$ silent | 313 | - | 269 | 266 | - | **306** | 225 | - | 182 | 250 | - | 232 | 214 |
| 4$^-$ TO3 | 480 | 468 | 471 | 455 | 534 | **487** | 518 | 513 | 483 | 557 | **505** | 542 | 496 |
| 4$^-$ LO1 | 182 | 173 | 172 | 162 | 111 | **180** | 146 | - | - | - | - | 134 | 115 |
| 4$^-$ LO2 | 472 | 453 | 439 | 434 | 412 | **468** | 397 | - | - | - | - | 395 | 214 |
| 4$^-$ LO3 | 677 | 738 | 683 | 657 | 607 | **717** | 677 | - | - | - | - | 657 | 650 |

4$^-$ : a (000), $t_{1u}$ (x,y,z); b(0.5,0.5,0.5), $t_{1u}$ (x,y,z); d $a_{2u}$ (z),$e_u$(x,y) ; 5$^-$ : d $e_u$(x,y). * experimental data are unknown

Table 3  Phonon frequencies (cm$^{-1}$) at X,M and R points of Brillouin zone

| k-point, site symmetry | symmetry | BaTiO$_3$, Present work | BaTiO$_3$, [5] | BaTiO$_3$, [35] | symmetry | BaZrO$_3$ | symmetry | BaHfO$_3$ |
|---|---|---|---|---|---|---|---|---|
| **X (0,0,0.5)** | 5$^-$ | 259i, 96i[9] | 189i | 84i | 5$^+$ | 95 | 5$^+$ | 99 |
| a (000) | 5$^+$ | 124 | 104 | 109 | 5$^-$ | 131 | 5$^-$ | 103 |
| t$_{1u}$ (x,y,z)   3$^-$5$^-$ | 1$^+$ | 162 | 152 | 147 | 1$^+$ | 140 | 1$^+$ | 144 |
|  | 5$^-$ | 216 | 195 | 204 | 3$^-$ | 229 | 3$^-$ | 179 |
| b (0.5 0.5 0.5) | 3$^-$ | 285 | 261 | 285 | 5$^-$ | 231 | 5$^-$ | 236 |
| t$_{1u}$(x,y,z)   1$^+$5$^+$ | 4$^-$ | 352 | 323 | 324 | 4$^-$ | 266 | 4$^-$ | 272 |
|  | 5$^+$ | 360 | 331 | 327 | 5$^+$ | 285 | 5$^+$ | 288 |
| d (000.5) | 5$^-$ | 445 | 423 | 439 | 3$^-$ | 466 | 3$^-$ | 433 |
| a$_{2u}$(z)   1$^+$5$^-$ | 3$^-$ | 557 | 519 | 518 | 5$^-$ | 492 | 5$^-$ | 517 |
| e$_u$(x,y)   3$^-$4$^-$5$^-$5$^+$ | 1$^+$ | 670 | 668 | 630 | 1$^+$ | 735 | 1$^+$ | 745 |
| **M(0,0.5,0.5)** | 3$^-$ | 241i, 9i[9] | 167i | 85i | 5$^+$ | 93 | 5$^+$ | 95 |
| a (000) | 5$^-$ | 117 | 105 | 111 | 2$^-$ | 94 | 2$^-$ | 98 |
| t$_{1u}$(x,y,z)  3$^-$5$^-$ | 2$^-$ | 125 | 104 | 109 | 3$^+$ | 124 | 3$^-$ | 119 |
|  | 3$^+$ | 239 | 209 | 194 | 3$^-$ | 153 | 3$^+$ | 128 |
| b (0.5 0.5 0.5) | 5$^-$ | 302 | 271 | 284 | 5$^-$ | 228 | 5$^-$ | 196 |
| t$_{1u}$(x,y,z)  2$^-$5$^-$ | 3$^-$ | 344 | 334 | 341 | 5$^+$ | 297 | 5$^+$ | 300 |
|  | 2$^+$ | 358 | 345 | 356 | 5$^-$ | 370 | 5$^-$ | 325 |
| d (000.5) | 5$^+$ | 374 | 356 | 364 | 4$^+$ | 431 | 4$^+$ | 435 |
| a$_{2u}$(z)   1$^+$2$^+$3$^-$ | 5$^-$ | 464 | 437 | 441 | 3$^-$ | 470 | 3$^-$ | 499 |
| e$_u$(x,y)   3$^+$4$^+$5$^+$5$^-$ | 4$^+$ | 494 | 458 | 460 | 2$^+$ | 548 | 2$^+$ | 578 |
|  | 1$^+$ | 737 | 686 | 722 | 1$^+$ | 789 | 1$^+$ | 796 |
| **R(0.5,0.5,0.5)** | 5$^+$ | 148,134[9] | 129 | 133 | 4$^+$ | 91 | 4$^+$ | 96 |
| a (000)   t$_{1u}$ (x,y,z)  4$^-$ | 4$^+$ | 217 | 183 | 167 | 5$^+$ | 114 | 5$^+$ | 118 |
| b (0.5 0.5 0.5)   t$_{1u}$ (x,y,z)   5$^+$ | 3$^+$ | 325 | 315 | 389 | 5$^-$ | 317 | 5$^-$ | 227 |
| d (000.5) | 5$^-$ | 413 | 388 | 399 | 5$^+$ | 389 | 5$^+$ | 392 |
| a$_{2u}$(z)   1$^+$3$^+$ | 5$^+$ | 450 | 416 | 417 | 3$^+$ | 538 | 3$^+$ | 570 |
| e$_u$(x,y)   4$^+$5$^+$ | 1$^+$ | 772 | 720 | 754 | 1$^+$ | 819 | 1$^+$ | 823 |

Mulliken atomic charges Q, given in Table 2, demonstrate high covalence in the transition metal-oxygen interaction. The calculation of atomic Born charges B was made with the experimental high frequency dielectric constant $\varepsilon^\infty$ =5.4 found for BaTiO$_3$ by extrapolating refractive index measurements at different wavelength to zero frequency [5] and $\varepsilon^\infty$ = 4.6 (found in [14,15]) for two other crystals. The numerical values of Born charges in BaTiO3 agree with those found in PW DFT calculations and the experimental estimates [35], see Table 2.

The results given in Table 3 demonstrate that the structural instability in BaTiO$_3$ is may be connected also with imaginary phonon frequencies at M and X points of BZ, but not at R point. This is in agreement with the results of DFT PW calculations [5,7,9] on phonons in BaTiO$_3$.

The experimentally observed stability of the cubic structure for BaZrO$_3$ and BaHfO$_3$ is in agreement with the results of the present work as no imaginary frequencies were obtained for these crystals both at the BZ symmetry points (see Table 3) and on the six symmetry directions under consideration.

Let us summarize the results obtained.

Pioneer first-principles LCAO computations on Ba(Ti,Zr,Hf)O$_3$ are performed within the hybrid Hartree-Fock-DFT exchange –correlation functional. We have clearly shown that LCAO calculations with the basis set optimization and use of hybrid PBE0 functional are able to predict correctly the structural stability and phonon properties in cubic perovskites under consideration.

The agreement between the experimental and calculated ground state properties and phonon frequencies is better in present work than that obtained in traditional PW DFT calculations.

The results as obtained in the present study for BaHfO$_3$ are very important for future experimental study of the vibrational spectra of this crystal as at present such studies are unknown.

Author is grateful to Prof. Manuel Cardona, Prof. Eugene Kotomin and Dr. Denis Gryaznov for helpful discussions, to Dr. A. Bandura and Maxim Losev for the support with the OPTBAS package.

* re1973@re1973.spb.edu